\documentclass{elsart}

\usepackage{graphicx}

\begin{document}
\begin{frontmatter}
\title{Electron-electron interaction in multiwall carbon nanotubes\thanksref{talk}}
\thanks[talk]{Expanded version ..}

\author{A.I. Romanenko, A.V. Okotrub, O.B. Anikeeva, L.G. Bulusheva, N.F. Yudanov}
\address{$^1$Institute of Inorganic Chemistry Siberian Branch of Russian Academy of Science,
Novosibirsk, Russia}
\author{C. Dong, Y. Ni}
\address{$^2$National Laboratory for Superconductivity, Institute of Physics Chinese Academy
of cience, Beijing, China}

\begin{abstract}
Magnetic susceptibility $\chi$ of pristine and brominated
arc-produced sample of multiwall carbon nanotubes was measured
from 4.2 to 400 K. An additional contribution $\Delta \chi(T)$  to
diamagnetic susceptibility $\chi(T)$ of carbon nanotubes was found
at T $<$ 50 K for both samples. It is shown that $\Delta \chi(T)$
are dominated by quantum correction to $\chi$ for interaction
electrons (interaction effects-IE). The IE shows a crossover from
two-dimensional to three-dimensional at $B$ = 5.5 T. The effective
interaction between electrons for interior layers of nanotubes are
repulsion and the electron-electron interaction $\lambda$$_c$ was
estimated to be $\lambda_c\sim $ 0.26.
\end{abstract}
\end{frontmatter}

From the time of discovery nanotubes one of the most important
problems is the possibility of a superconducting state in them. In
a series of experiments were observed the effects which indicate
on its existence. There is supercurrents through single-walled
carbon nanotubes ~\cite {Kasu99}, persistent currents and magnetic
flux trapping ~\cite {Tseb99}. As is known the nature of a
superconducting state is the electron-electron interaction (
Namely - attraction between electrons). On the other hand
electron-electron interaction is exhibited in electronic transport
properties of conductors in normal state - so-called quantum
interference effects - interaction effects (IE) ~ \cite
{Lee85,Al't83}. IE are connected with the correction to density of
states of conduction electrons in a results of quantum
interferences of electrons at their diffuse motion in random
conductors. But in such systems the one-particle processes,
so-called weak localization (WL) ~ \cite {Lee85} and weak
antilocalization (WAL) ~ \cite {Hika80}, always accompany with IE.

The observation of IE corrections to $\chi$ is very important for
partition of including of WL, WAL, and IE to different physical properties. From all these
corrections only IE contributes to $\chi$.

For observation of IE corrections to $ \chi $ it is necessary to
divide the contributions connecting with IE and much more on
quantity a magnetic susceptibility of sample, and exclude the
contribution of paramagnetic impurities. Only in separate cases it
is possible. Earlier, with the using of relaxation processes in
Mo$_ 2$S $ _ 3 $ ~\cite {Roma85a}, we changed the contribution
connected with IE correction to $ \chi $ by quenching of
high-temperature metastable state of a sample. In a results, with
the using of difference contribution to $ \chi $ in equilibrium
and metastable states, we received the IE correction to $ \chi $
in the pure state ~\cite {Roma85b}. In this work, with the using
of chemical modification of a sample (brominated), we singled out
IE correction to $ \chi $ in laminated structures based on
multi-layer carbon nanotubes (LS of MWNT).

The material which contained MWNTs was synthesized using a set-up
for arc discharge graphite evaporation, which was described
elsewhere~\cite{Okot96,Okot95}. The arc was maintained with a
voltage of 35 V and a current of 1000 A for 15-20 minutes in
helium atmosphere of 800 Torr. A nanotube content in the inner
part of carbon deposit grown on the cathode was estimated by
transmission electron microscopy to reach about
$80\%$~\cite{Okot00}. Tubes have from 2 to 30 shells with an outer
diameter of 60-150 {\AA }. Scanning electron microscopy (SEM)
revealed a predominant orientation of MWNTs was perpendicular to
the deposit growth axis (Fig. 1).

\begin{figure}
\begin{center}
\includegraphics[scale=0.8]{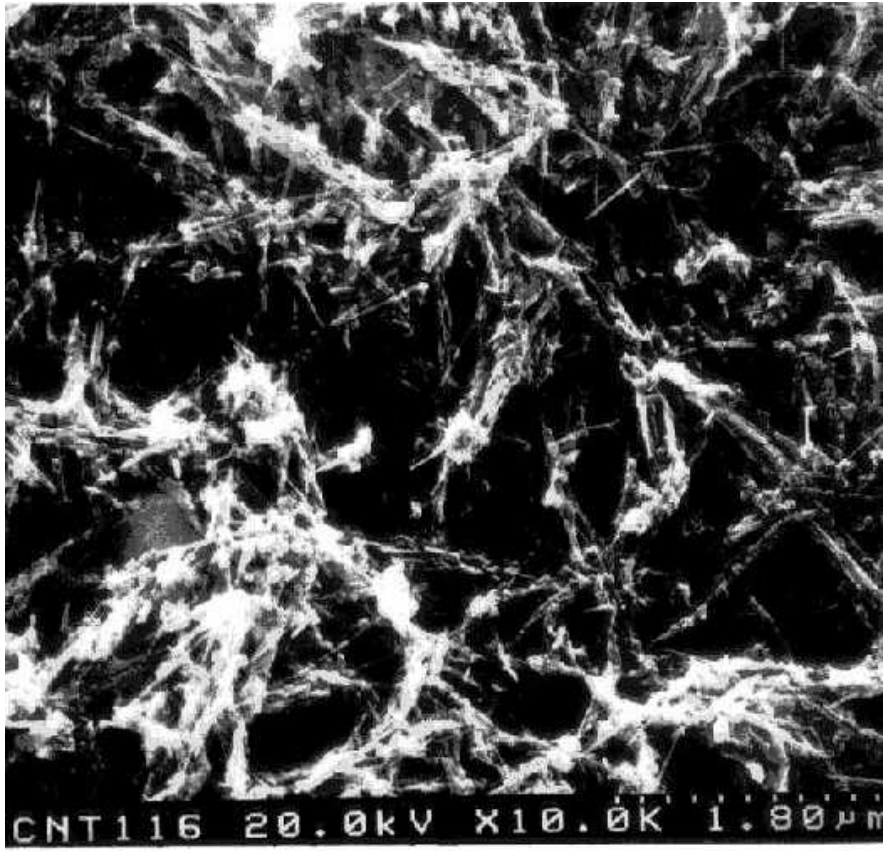}
\end{center}
\caption{The SEM micrograph of prestine macro-samples LS of MWNT.
The deposit axis was perpendicular to picture.} \label{ret}
\end{figure}

\newpage

The brominated material was prepared by exposure of the pristine
one to Br$_2$ vapors during 7 days. Its composition CBr$_{0.06} $
was determined by x-ray photoelectron spectroscopy.

For magnetic measurements a cylindrical sample of diameter 5mm and
of length 10 mm was cut out from the pristine or brominated
material so  that its axis was coincident with the deposit one. As
a result, the magnetic field was perpendicular to most of carbon
nanotubes in the sample. The weight of samples was about 0.3
grams. The temperature dependence of magnetic susceptibility $
\chi $ for the samples was measured from 4.5 to 400 K in a field
of 0.01, 0.5, and 5.5 T by using a model MPMS-5 SQUID (QUANTUM
DESIGN, USA).

According to experimental and theoretical data, the basic
contribution in $ \chi $ of quasi-two-dimensional graphite (QTDG),
including MWNTs, gives orbital magnetic Susceptibility $ \chi _
{or} $  connected with extrinsic carriers (EC) ~ \cite
{Koto97,Koto87,Koto91}.

Figure 2(a) presents the magnetic susceptibility $\chi $ of
pristine sample as a function of temperature. The observed
behavior was similar to the previously reported measurements on
the samples of MWNTs~ \cite{Koto97,Koto87,Koto91,Here94,Tsui00}.

\begin{figure}
\begin{center}
\includegraphics[scale=0.29]{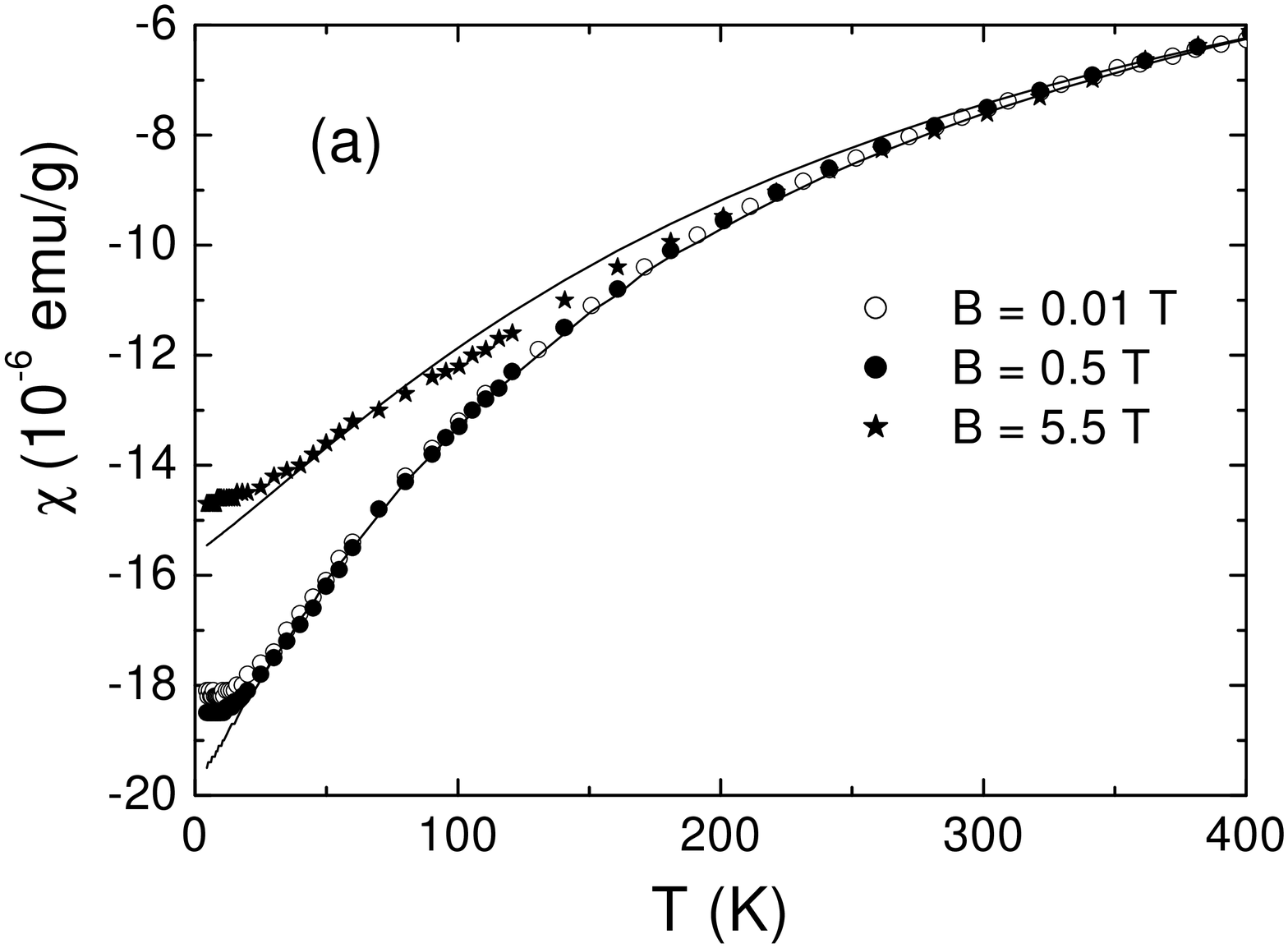}
\end{center}
\label{ret}
\end{figure}
\begin{figure}
\begin{center}
\includegraphics[scale=0.28]{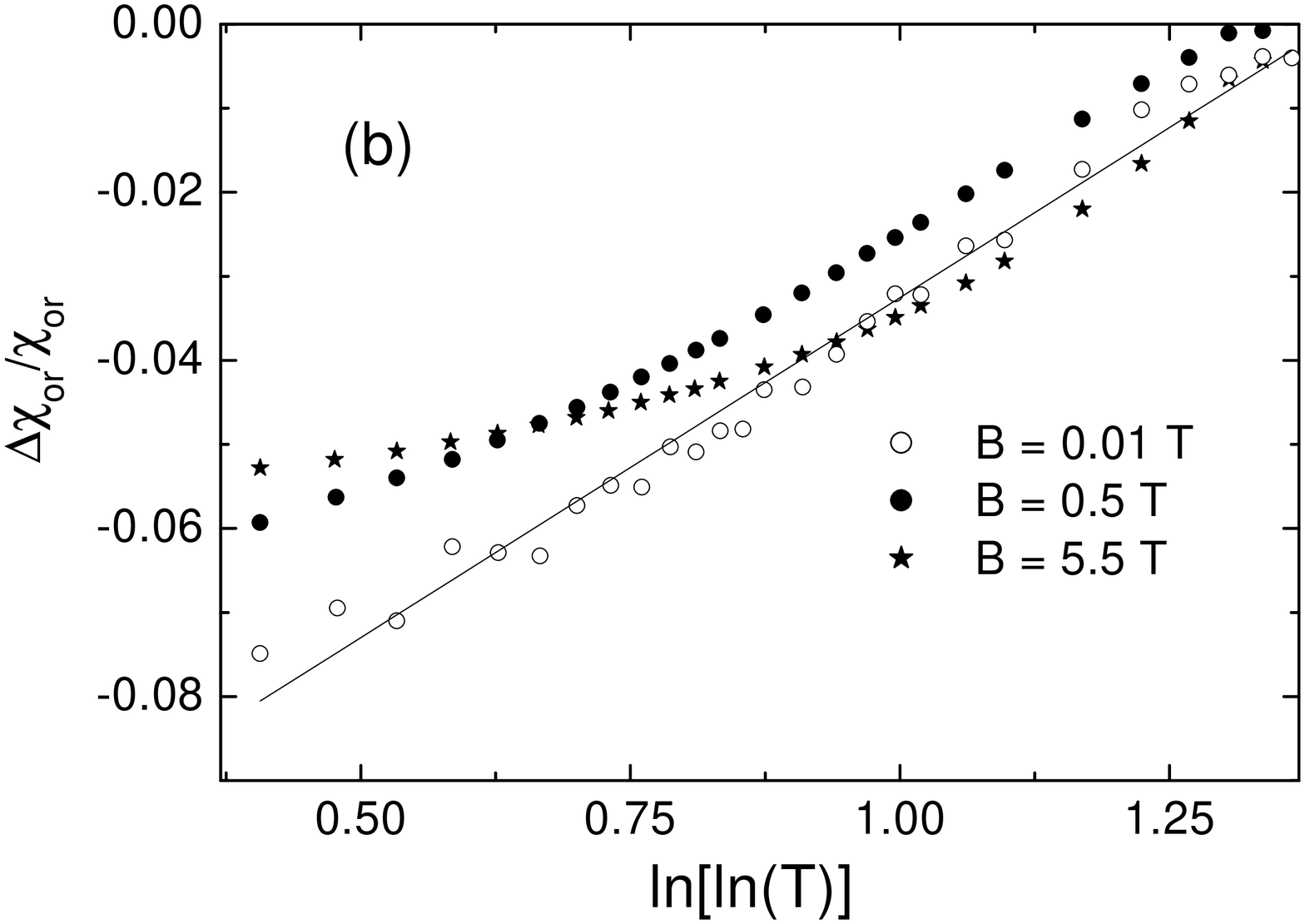}
\end{center}
\label{ret}
\end{figure}
\begin{figure}
\begin{center}
\includegraphics[scale=0.28]{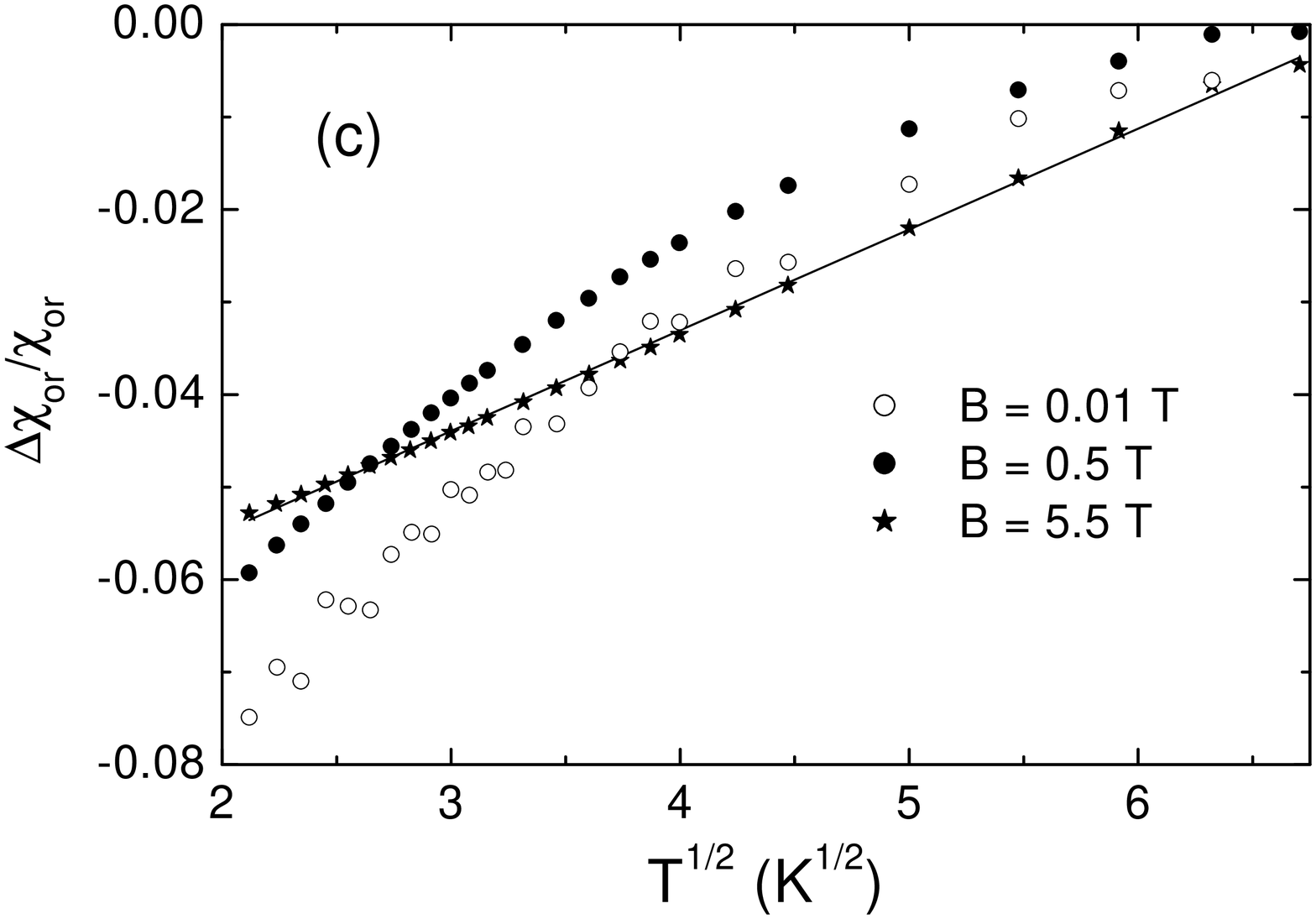}
\end{center}
\caption{The temperature dependence of magnetic susceptibility
$\chi (T)$ ({\bf a}) and $\Delta \chi_{or} (T)/\chi _{or}(T)$ =
[$\chi (T) -\chi _{or}(T)]/\chi _{or}(T)$ [({\bf b}) and ({\bf
c})] for prestine sample. The solid lines are fits: for ({\bf a})
by Eq. (1) in interval 50 - 400 K with parameters; for curve
($\circ$) , $\gamma _0$ = 1.6 eV, $T_0$ = 215 K, $\delta$  = 159
K; for ($\bullet$)  , $\gamma _0$ = 1.6 eV, $T_0$ = 215 K,
$\delta$  = 159 K; for ($\star$)  , $\gamma _0$ = 1.7 eV, $T_0$ =
327 K, $\delta$  = 210 K; by Eq. (2) and Eq. (3) for ({\bf b}) and
({\bf c}) respectively in interval 4.5 - 45 K with parameters
$T_c$ = 10000 K , $l_{el}/a$ = 0.15.} \label{ret}
\end{figure}

\newpage

Available models well reproduce the temperature dependence of
magnetics susceptibility for MWNTs only at T $>$ 50 K~ \cite
{Koto97,Koto87,Koto91,Ajik93,Lu95}. In the low-temperature region
the experimental data deviate from the theoretical ones that are
usually attributed to the paramagnetic impurities contribution. To
analyze this anomalous part of the magnetic susceptibility $ \chi
$ in detail, that was a goal of the our work, it was necessary
 to select its high-temperature portion previously. According to
theoretical consideration the magnetic susceptibility $\chi $ of
quasi-two-dimensional graphite (QTDG) is generally contributed by
two components: diamagnetic susceptibility $\chi_D $ ~ \cite
{Koto97,Koto87,Koto91,Ajik93,Lu95} and paramagnetic spin
susceptibility $\chi_s $. The amount of metallic impurities in the
samples under investigation was detected by spectrographic
analysis and was less than $10^{-5}$. Furthermore, the signal
corresponding to the unpaired spins of transition metals was
absents in the EPS spectrum of samples (detection limit equal to
$10^{-6}$). Thus the paramagnetic spin contribution to the
magnetic susceptibility $ \chi $ of the measured materials was
negligible. Hence, the $\chi$ of MWNT sample have been determined
by the component $\chi_D $~\cite{Koto97}

\begin{equation}
\chi _{or}(T) = -\frac{5.45\times 10^{-3}\gamma_0^2}{(T+\delta )[2+exp(\eta )+exp(-\eta )]} ,
\end{equation}

where $\gamma_0$ is the band parameter for two-dimensional case,
$\delta $ is the additional temperature formally taking into
account "smearing" the density of states due to electron
nonthermal scattering by structure defects, $\eta $ =
$E_F/k_B(T+\delta )$ represents reduced Fermi level ($E_F$), $k_B$
is the Boltzmann constant. Using an electrical neutrality equation
in the 2D graphite model ~\cite{Koto97} $\eta $ can be derived by
$\eta $ = sgn($\eta _0$)[0.006$\eta _0^4$ - 0.0958$\eta _0^3$ +
0.532$\eta _0^2$ - 0.08$\eta_0$] ~\cite{Koto87} with an accuracy
no less then $1\%$. The $\eta_0$ is determined by $\eta_0 $ =
$T_0/(T+\delta )$, where $T_0$ being degeneracy temperature of
extrinsic carriers (EC) depends on its concentration $n_0$ only.
The value of $\delta $ can be estimated independently
~\cite{Koto91} as $\delta $ = $\hbar $/$\pi k_B\tau _0$ ,
 where $\hbar $ is the Planck constant, $\tau _0$  is a relaxation time of the carrier
nonthermally scattered by defects~\cite{Koto91}. Generally, the
number of EC in QTDG is equal to that of scattering centers and
$\delta$ depends only on $T_0$, i.e. $\delta  = T_0/r$, where $r$
is determined by scattering efficiency ~\cite{Koto87}. These
parameters were chosen to give the best fit of the experimental
data (Fig. 2a). At high field ($B  >  $ 1  T) the magnetic
susceptibility $\chi (T)$ decreases in all interval of
temperature. In this field region the magnetic
length~\cite{Lee85,Al't83} $l_B$ = $(\hbar c/2eB)^{1/2}$ is much
less than a tube length. Therefore, susceptibility probes only
small local areas of the graphite plane, and is expected to be the
geometrical averaged of that of rolled-up sheets of graphite. At
low fields the magnetic length is larger than the dimension of
most tubes in the sample, and this geometrical correction may be
neglected~\cite{Here94}.

The data in Fig. 2 show that at T$ < $50 K there is an additional
contribution $\Delta \chi _{or}(T)=\chi (T) -\chi _{or}(T)$ to
$\chi (T)$. According to theoretical
calculations~\cite{Lee85,Al't83} only electron-electron
interactions can contribute to magnetic susceptibility, which may
be divided into two parts. The first part $\Delta \chi _{or}$
comes from correction to orbital susceptibility $\chi_{or}$. The
other part $\Delta \chi _s$ is associated with the correction to
spin susceptibility $\chi_s$. The latter contribution is
negligible because the carbonaceous material used in sample
preparation was indicated by spectrographic analysis and these
material contained very small amount of magnetic impurities (less
than detection limit). The dominated part $\Delta \chi _{or}$,
divided on $\chi _{or}$, was described by~\cite{Lee85,Al't83}

\begin{equation}
 \frac{\Delta \chi_{or} (T)}{\chi _{or}(T)}=
-\frac{\frac{4}{3}(\frac{l_{el}}{h})ln[ln(\frac{T_c}{T})]}
{ln(\frac{k_B T_c\tau _{el}}{\hbar })}      ,(d = 2)  ,
\end{equation}
\begin{equation}
 \frac{\Delta \chi_{or} (T)}{\chi _{or}(T)}=
-\frac{2(\frac{\pi }{6})\xi (\frac{1}{2})
(\frac{k_B T\tau _{el}}{\hbar })^{1/2}}{ln(\frac{T_c}{T})}      ,(d = 3)  ,
\end{equation}

where value of $\xi (\frac{1}{2})$ $\sim $ 1, $l_{el}$ is the
electron mean free path; $\tau _{el}$ represents the elastic
relaxation time, which is about  10$^{-13}$ sec for MWNT
~\cite{Baxe97}; $h $ is the thickness of layer in two-dimensional
case; $d $ denoted a system dimensionality; $T_c$ = $\theta
_Dexp(\lambda _c^{-1}$), where $\theta _D$ is the Debye
temperature and $\lambda _c$ is the constant described the
electron-electron interaction in Cooper canal ($\lambda _c  >  0 $
in a case of electron repulsion). The dependence in Eq. (2) is
determined by $ln[ln(\frac{T_c}{T})]$ term because at low
temperature, in the disordered systems, $\tau _{el}$ is the
temperature independent while all other terms are constants. The
dependence in Eq. (3) is governed by $\ T^{1/2}$ term as $T_c \gg
T $ and, therefore, $ln(\frac{T_c}{T})$ is constant relative to $\
T^{1/2}$.

The additional contribution to $\chi (T)$ as a function of
$ln[ln(\frac{T_c}{T})]$ and $\ T^{1/2}$ is presented in Fig. 2(b)
and Fig. 2(c). The $\triangle \chi_{or} (T)/\chi _{or}(T)$ clearly
shows the dependence given by Eq. (2) at low magnetic field and
one given by Eq. (3) at high magnetic field, while at $B$ = 0.5 T
the temperature behavior of $\triangle \chi_{or} (T)/\chi
_{or}(T)$ is deviated from these two cases. As seen from Fig. 2,
an absolute value of $\triangle \chi_{or} (T)/\chi _{or}(T)$ at
all magnetic fields applied to the pristine sample increases with
decreasing temperature that has been predicted for IE in the
systems characterized by electron-electron repulsion
~\cite{Lee85,Al't83}. Hence, at $B$ = 5.5 T a transfer from
two-dimensional IE correction to three-dimensional one takes
place. At lower magnetic field the interaction length $L_I(T)$ =
$(\hbar D/k_BT)^{1/2}$ is much less than the magnetic length $l_B
= (\hbar c/2eB)^{1/2}$, which in turn becomes dominant at high
field. An estimation of the characteristic lengths gave
respectively the value of $L_I(4.2K)$ = 130 {\AA } (taking into
account that the diffusion constant $D$ = 1
cm$^2$/s~\cite{Baxe97}) and the value of $l_B$ = 100 {\AA } at $B$
= 5.5 T.

Taking into account experimentally apparent crossover (Fig. 2)
from two-dimensional (equation 2) to three-dimensional (equation
3) of temperature dependence of $ \triangle \chi _ {or} (T) /\chi
_ {or} (T) $ we concluded that effective diameter of nanotubes was
in an interval between 100 {\AA} and 130 {\AA}.

Partitioning of the contributions $ \chi _ {or} (T) $ and $
\triangle \chi _ {or} (T) /\chi _ {or} (T) $ can be carried out in
a result of changing of extrinsic carriers by the chemical
modification of the prestine sample. In this case the $ \chi _
{or} (T) $ should be changed ~\cite {Koto97} and the $ \triangle
\chi _ {or} (T) /\chi _ {or} (T) $ remain the invariable ~\cite
{Lee85,Al't83}.

\begin{figure}
\begin{center}
\includegraphics[scale=0.28]{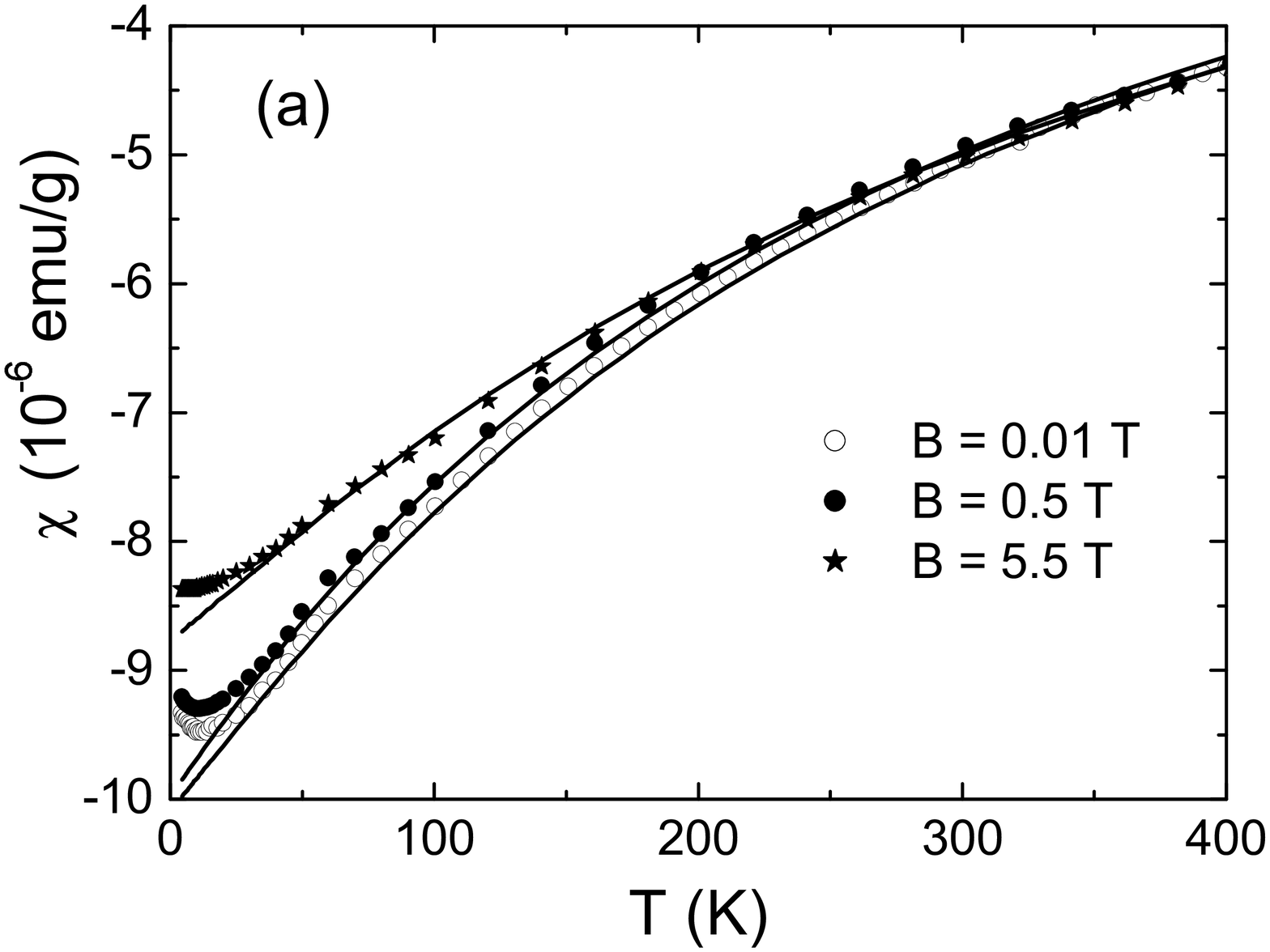}
\end{center}
\label{ret}
\end{figure}
\begin{figure}
\begin{center}
\includegraphics[scale=0.26]{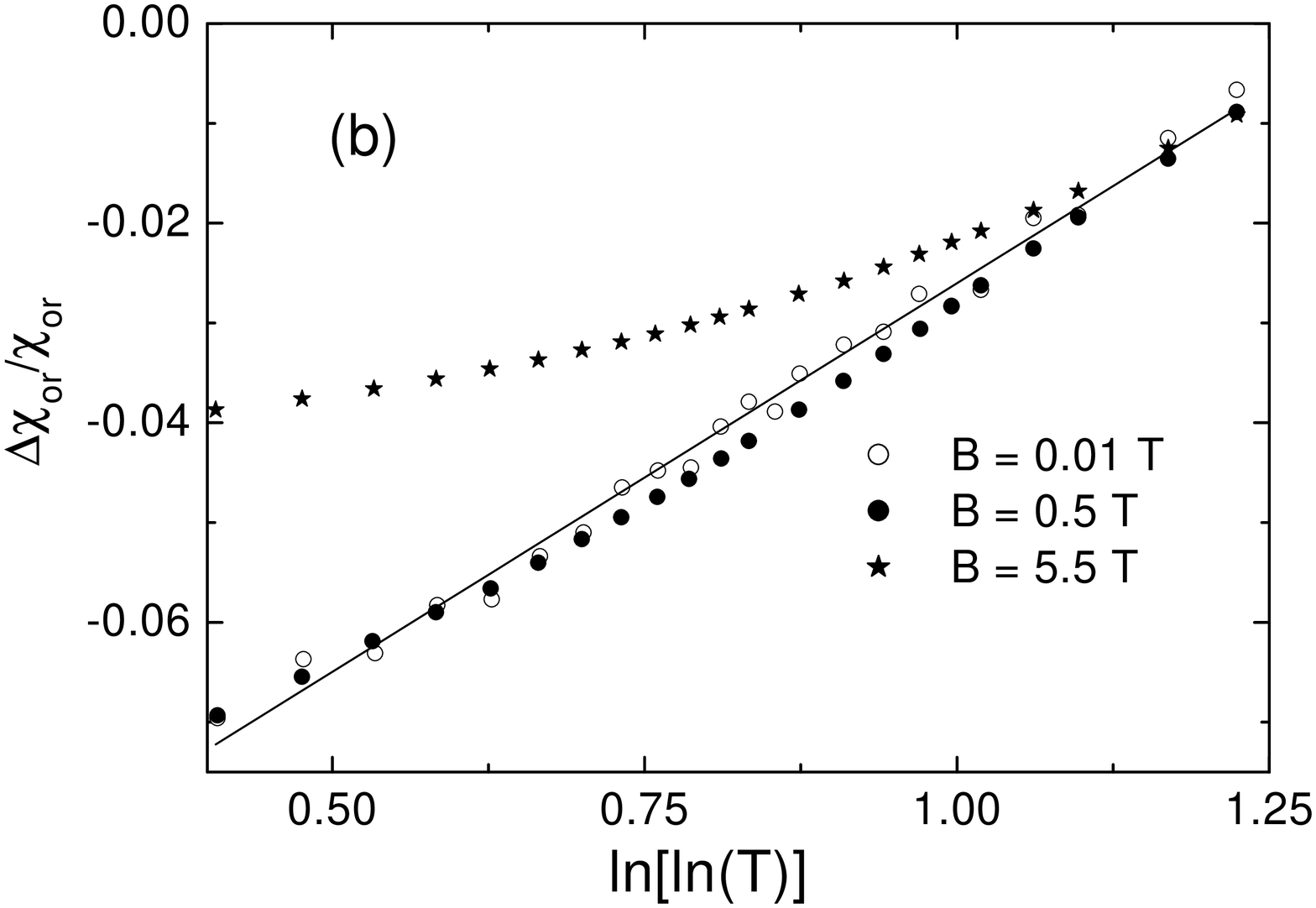}
\end{center}
\label{ret}
\end{figure}
\begin{figure}
\begin{center}
\includegraphics[scale=0.26]{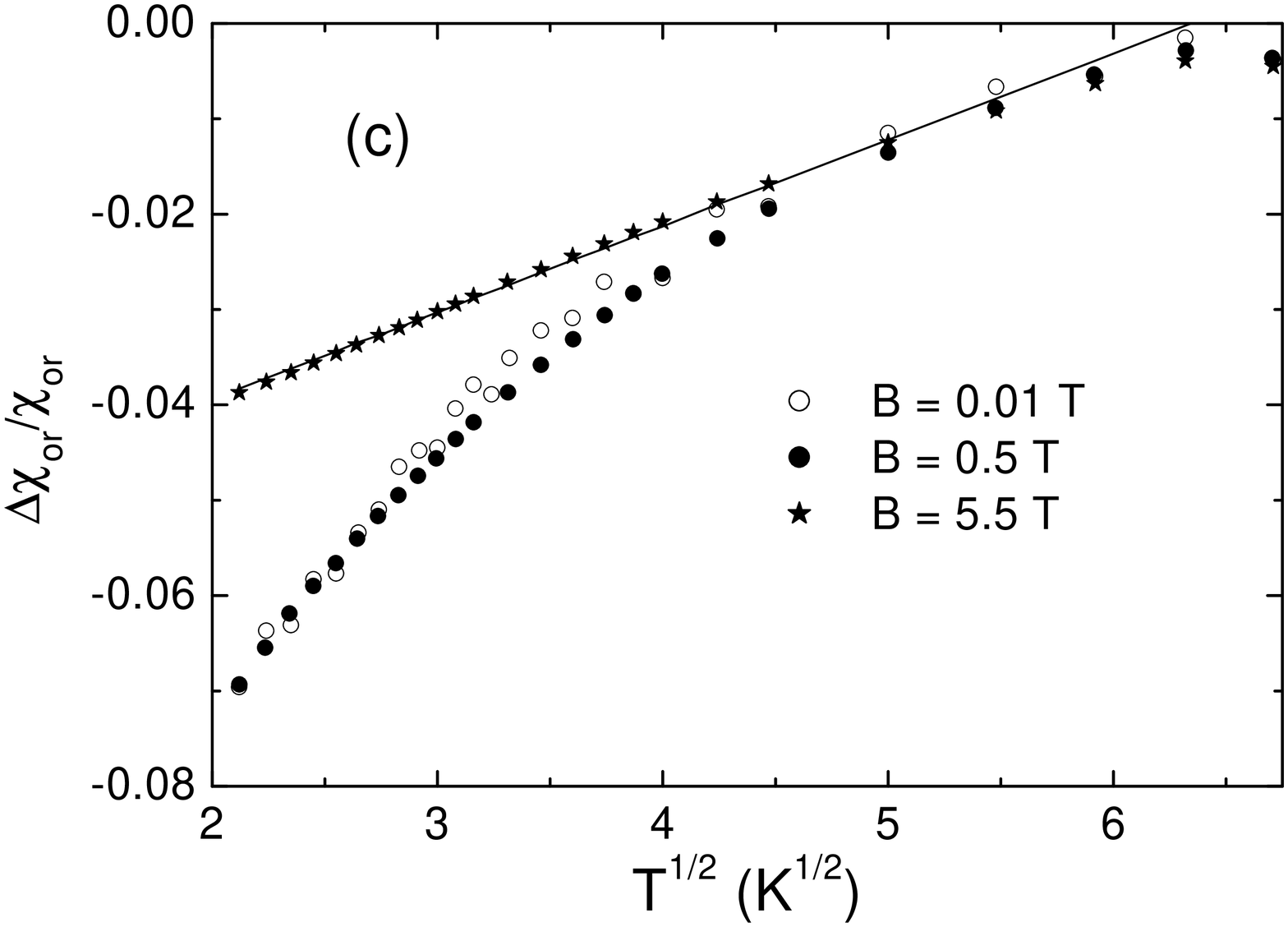}
\end{center}
\caption{The temperature dependence of magnetic susceptibility
$\chi (T)$ ({\bf a}) and $\Delta \chi_{or} (T)/\chi _{or}(T)$ =
[$\chi (T) -\chi _{or}(T)]/\chi _{or}(T)$ [({\bf b}) and ({\bf
c})] for bromineted sample. The solid lines are fits: for ({\bf
a}) by Eq. (1) in interval 50 - 400 K with parameters; for curve
($\circ$) , $\gamma _0$ = 1.4 eV, $T_0$ = 340 K, $\delta$  = 252
K; for ($\bullet$)  , $\gamma _0$ = 1.4 eV, $T_0$ = 300 K,
$\delta$  = 273 K; for ($\star$)  , $\gamma _0$ = 1.5 eV, $T_0$ =
435 K, $\delta$  = 325 K; by Eq. (2) and Eq. (3) for ({\bf b}) and
({\bf c}) respectively in interval 4.5 - 45 K with parameters
$T_c$ = 10000 K , $l_{el}/a$ = 0.15.} \label{ret}
\end{figure}

\newpage

The intercalation of single wall carbon nanotubes (SWNT) by
bromine lied to increase of $n_0$~\cite{Rao97}. We supposed that
in MWNT may be similarly situation and used brominated MWNT for
our investigations. Figure 3 show $\chi (T)$ for the brominated
sample. The $T_0$ and $\delta $ for these sample are shown in
figure 3 caption. We estimated $n_0$ at low temperature and low
field in framework of theory of QTDG~\cite{Koto91} $n_0 =
4(k_BT_0)^2/(3\pi a^2\gamma _0^2)$, where $a$ = 0.246 nm  -
lattice parameter in layer. These estimations gives: $n_{0ini}$
$\sim $ 3$\times$10$^{10}$ cm$^{-2}$ - for prestine sample;
$n_{0Br} \sim $  10$^{11}$ cm$^{-2}$ - for brominated sample. The
$n_0$ increases in brominated sample about 3 times.  According to
the Drude formula, the conductivity is proportional to $n_0$ and
$\tau _{el}$. We measured the conductivity of pristine and
brominated samples and find that conductivity increase in 3 times
in a result of bromination. Taking into account that $n_0$ also
increase by about 3 times in a result of bromination we conclude
that $\tau _{el}$ practically did not change during bromination.
The $\Delta \chi_{or} (T)/\chi _{or}(T)$ don't change during
intercalation [Fig. 3(b) and Fig. 3(c)]. This result indicates
that $\Delta \chi_{or} (T)/\chi _{or}(T)$ doesn't connected with
n$_0$, and connected with the IE correction to $\chi$. From
experimental data $\Delta \chi_{or} (4.5K)/\chi _{or}(4.5K)$ =
0.027 for three-dimensional IE correction to $\chi$ we estimated
the $T_c$ from Eq. (3) and fined $T_c$ = 5$\times $10$^4$ K. We
estimated the $\lambda _c$ from $T_c = \theta _Dexp(\lambda
_c^{-1})$ with Debye temperature for carbon
nanotubes~\cite{Bene96} $\theta _D$ = 1000 K and obtained $\lambda
_c \sim $ 0.26. From experimental data $\Delta \chi_{or}
(4.5K)/\chi _{or}(4.5K)$ = 0.07 for two-dimensional IE correction
to $\chi$ we estimated the $l_{el}/h$ from Eq. (2) and fined
$l_{el}/h \sim $ 0.15. If $h \sim d_m$ and 100 {\AA } $\leq d_m
\leq$ 130 {\AA } so  20 {\AA } $\leq l_{el} \leq $ 15 {\AA }. This
estimation in a quite good agreement with estimation $l_{el} =
(D\tau _{el})^{1/2} \sim $ 30 {\AA } for so crude estimation.

In summary, we have investigated the additional contribution to
temperature dependence of orbital magnetic susceptibility $\Delta
\chi_{or} (T)/\chi _{or}(T)$ of lamination structure of multiwall
carbon nanotubes at $T < $ 50 K. It is shown that $\Delta
\chi_{or} (T)/\chi _{or}(T)$ is connected with quantum correction
to magnetic susceptibility for interaction electron. At low field
this correction is two-dimensional. At $B$ = 5.5 T was observed
three-dimensional correction to magnetic susceptibility. This
crossover from $2d$ to $3d$ behavior of $\Delta \chi_{or} (T)/\chi
_{or}(T)$ is connected with decreasing of magnetic length up to
value less then typical mean diameter of nanotubes. It is shown
that brominated of samples lead to increasing of extrinsic
carriers $n_0$ from $n_{0ini} \sim $ 3$\times $10$^{10}$ cm$^{-2}$
for prestine sample up to $n_{0Br} \sim $ 10$^{11}$ cm$^{-2}$ for
brominated samples. But $\Delta \chi_{or} (T)/\chi _{or}(T)$ did
not changed when $n_0$ increases, which is in full agreement with
theoretical predictions. From $\Delta \chi_{or} (4.5K)/\chi
_{or}(4.5K)$, we estimated the constant of electron-electron
interaction $\lambda _c \sim $ 0.26. This interaction is repulsion
for interior layers, which give the domination contribution to
$\Delta \chi_{or} (T)/\chi _{or}(T)$ as a integration value.

It is necessary to note, that the correction to a magnetic
susceptibility observation by us, and, accordingly estimation of a
constant of electron-electron interaction it is integrated values
average on all stratums in structural nanotubes. Therefore the
made conclusion about repulsion character of interaction between
electrons does not eliminate opportunity of an attraction between
electrons in high layer nanotubes, which contribution in magnetic
susceptibility is small in comparison with sum by the contribution
of all remaining stratums of a tube.

\section{Acknowledgements}
The authors thank Ms. Chaoying WANG for the SEM analysis of the
samples, and Dr. V.A. Nadolinny for the EPR measurements. The work
was supported by Lu Jiaxi grant for international joint research
from Chinese Academy of Sciences, and Russian scientific and
technical program "Fullerenes and atomic clusters" (Projects No
5-1-98), INTAS (Grant Nos 97-1700, 00-237), Russian Foundation of Basic Research
(Grants No: 00-02-17987; 00-03-32510; 00-03-32463; 01-02-0650),
and Interdisciplinary Integral Program of Siberian Branch of
Russian Academy of Science (Grant No 61).

\end{document}